\documentclass[graybox, envcountchap]{svmult}

\usepackage{mathptmx}        
\usepackage{amsmath}
\usepackage{amssymb}
\usepackage{color}
\usepackage{helvet}          
\usepackage{courier}         
\usepackage{dirtree}

\usepackage{makeidx}      
\usepackage{graphicx}        
\usepackage{subfig}

\usepackage{multicol}        
\usepackage[bottom]{footmisc}

\usepackage{hyperref}        
\hypersetup{colorlinks=true,urlcolor=blue}

\usepackage[misc]{ifsym}
\usepackage{tikz}
\usetikzlibrary{positioning, arrows.meta, shapes.geometric, patterns, decorations.pathreplacing, calc}
\usepackage{pgfplots}
\pgfplotsset{compat=1.18}

\makeindex        

\begin{document}
\title{Time Series Analysis in Machine Learning}
\titlerunning{Time Series Analysis in Machine Learning}
\author{Antonio Pagliaro and Anna Anzalone}
\authorrunning{A. Pagliaro and A. Anzalone}
\institute{Antonio Pagliaro (\Letter) and Anna Anzalone \at INAF IASF Palermo, Via Ugo La Malfa 153, Palermo, I-90146, Italy \\}
%
%
\maketitle

\abstract{Time series analysis is a fundamental component of machine learning, especially in astrophysics and cosmology where temporal data abound. This chapter provides a pedagogical review of time series analysis techniques from a machine learning perspective. We cover the basic concepts of time series (stationarity, autocorrelation, seasonality), classical statistical models (autoregressive, moving average, ARIMA, exponential smoothing, state-space models), and modern machine learning approaches. In particular, we discuss how traditional statistical methods lay the groundwork, and then explore machine learning methods for time series, including feature-based regression, tree-based ensemble methods, hidden Markov models, Gaussian processes, and deep learning models (recurrent neural networks, convolutional networks, transformers). Throughout, we illustrate with examples drawn from multiple domains (e.g. astronomy, weather forecasting, finance) to emphasize common principles. The goal is to equip readers with both the theoretical understanding and practical context to apply machine learning techniques for time series analysis in their research.}

\section{Introduction}
A \emph{time series} is a sequence of observations indexed by time, typically recorded at regular intervals. Time series analysis aims to extract meaningful statistics and patterns from temporal data and to forecast future values based on past observations. This is relevant across a wide range of disciplines: economics (e.g.\ stock prices, GDP), environmental science (climate records), engineering (sensor readings), medicine (physiological signals), and many others. In astrophysics and cosmology, time series are ubiquitous -- for example, the light intensity of a variable star measured over many nights forms a time series known as a \emph{light curve}. The rapid growth of data from time-domain surveys in astronomy has made time series analysis increasingly important: projects like the Zwicky Transient Facility (ZTF) \cite{Bellm2019} generate over $10^5$ event alerts per night, and the Vera Rubin Observatory's Legacy Survey of Space and Time (LSST) \cite{Ivezic2019} is expected to produce an order of magnitude more. These data streams, capturing flux changes or transient events, demand robust and scalable analysis techniques. Machine learning (ML) has become indispensable in this context, enabling automated classification, pattern discovery, and anomaly detection in massive datasets where traditional methods alone are insufficient. ML algorithms now play a key role in time-domain astronomy, from identifying exoplanet transits and classifying variable stars to detecting supernovae and characterizing gravitational-wave signals.

\begin{figure}[t]
\centering
\begin{tikzpicture}
\pgfplotsset{
    every axis/.append style={
        width=10cm,
        xmin=0, xmax=6.28,
        tick label style={font=\scriptsize},
        ylabel style={font=\small, at={(axis description cs:-0.08,0.5)}, anchor=center},
        no markers, thick
    }
}
\begin{axis}[
    name=ax1, height=2.8cm,
    ymin=-1, ymax=6,
    ytick={0,3,6},
    ylabel={Observed},
    xticklabels={},
]
\addplot[black, domain=0:6.28, samples=200]
    {0.5*x + 1.5*sin(deg(2*x))
     + 0.12*sin(deg(5.7*x + 1.2)) + 0.09*sin(deg(9.3*x + 2.7))
     + 0.08*sin(deg(13.1*x + 0.5)) + 0.06*sin(deg(17.4*x + 3.1))};
\end{axis}
\begin{axis}[
    name=ax2, at={(ax1.below south west)}, anchor=north west,
    yshift=-0.5cm, height=2.5cm,
    ymin=-0.5, ymax=4,
    ytick={0,2,4},
    ylabel={Trend},
    xticklabels={},
]
\addplot[blue, domain=0:6.28, samples=40] {0.5*x};
\end{axis}
\begin{axis}[
    name=ax3, at={(ax2.below south west)}, anchor=north west,
    yshift=-0.5cm, height=2.5cm,
    ymin=-2, ymax=2,
    ytick={-2,0,2},
    ylabel={Seasonal},
    xticklabels={},
]
\addplot[red!70!black, domain=0:6.28, samples=80] {1.5*sin(deg(2*x))};
\end{axis}
\begin{axis}[
    name=ax4, at={(ax3.below south west)}, anchor=north west,
    yshift=-0.5cm, height=2.5cm,
    ymin=-0.5, ymax=0.5,
    ytick={-0.4,0,0.4},
    ylabel={Residual},
    xlabel={Time}, xlabel style={font=\small},
]
\addplot[gray, domain=0:6.28, samples=200]
    {0.12*sin(deg(5.7*x + 1.2)) + 0.09*sin(deg(9.3*x + 2.7))
     + 0.08*sin(deg(13.1*x + 0.5)) + 0.06*sin(deg(17.4*x + 3.1))};
\end{axis}
\end{tikzpicture}
\caption{Illustration of additive time series decomposition. The observed series (a) is expressed as the sum of a linear trend (b), a periodic seasonal component (c), and a residual (d). In real data the residual is stochastic and irregular; here it is approximated by a schematic residual for illustration. Decomposition is a fundamental preprocessing step that reveals the underlying structure of a series.}
\label{fig:decomposition}
\end{figure}
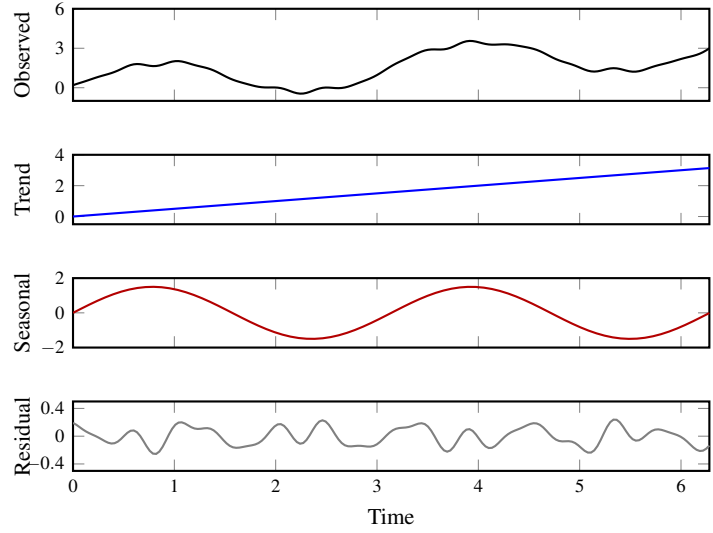

Time series analysis differs from standard data analysis because observations are not independent over time – instead, successive data points have temporal dependencies. The order and spacing of data points matter, and concepts like autocorrelation (correlation between past and future values) are fundamental. A crucial property in time series is \textit{stationarity}. Informally, a stationary time series is one whose statistical properties do not change over time. More precisely, a process $\{y_t\}$ is \emph{weakly stationary} (or covariance-stationary) if its mean $\mathbb{E}[y_t] = \mu$ is constant, its variance $\text{Var}(y_t) = \sigma^2$ is finite and constant, and its autocovariance $\text{Cov}(y_t, y_{t+k})$ depends only on the lag $k$, not on $t$ \cite{Hyndman2021}. In practice, real-world series often exhibit trends, seasonal patterns, or evolving variance that violate stationarity. For example, a raw light curve might brighten and dim periodically (seasonality) or gradually fade over time (trend). Many time series methods assume stationarity, so a common preprocessing step is to difference or detrend the data until it is approximately stationary. For instance, taking the daily change (first difference) of a stock price yields a more stationary series even if the original price had a trend. Seasonality (repeating patterns at fixed intervals such as daily or yearly cycles) can be addressed by seasonal differencing or by including seasonal terms in the model. A useful conceptual framework is \textit{time series decomposition}, which expresses a series as the sum (or product) of three components: a long-term \textit{trend}, a periodic \textit{seasonal} component, and a stochastic \textit{residual}. Classical decomposition and the more robust STL (Seasonal and Trend decomposition using Loess) method \cite{Cleveland1990} are widely used preprocessing steps that help analysts understand the structure of a series and isolate the component of interest before modeling (see Fig.~\ref{fig:decomposition} for an illustration).

Two fundamental diagnostic tools for understanding temporal dependencies are the \textit{Autocorrelation Function} (ACF) and the \textit{Partial Autocorrelation Function} (PACF). Intuitively, the ACF answers the question: ``how correlated is the series with a copy of itself shifted by $k$ time steps?'' Formally, the ACF at lag $k$ measures the linear correlation between $y_t$ and $y_{t-k}$:
\begin{equation}
\rho_k = \frac{\text{Cov}(y_t,\, y_{t-k})}{\text{Var}(y_t)} \,.
\label{eq:acf}
\end{equation}
The PACF at lag $k$ measures the correlation between $y_t$ and $y_{t-k}$ after removing the linear effects of intermediate lags $y_{t-1}, \ldots, y_{t-k+1}$. Concretely, it is the last coefficient $\phi_{kk}$ in the Yule--Walker fit of an autoregressive model of order $k$ (see Section~\ref{sec:arma-models}) to the series,
\begin{equation}
y_t \;=\; \phi_{k1}\, y_{t-1} + \phi_{k2}\, y_{t-2} + \cdots + \phi_{kk}\, y_{t-k} + \varepsilon_t\,.
\label{eq:pacf}
\end{equation}
Plotting the ACF and PACF of a series is a standard first step in time series analysis: the patterns they exhibit guide the choice of model. For instance, an ACF that decays exponentially with a PACF that cuts off after lag $p$ suggests an autoregressive AR($p$) process, while an ACF that cuts off after lag $q$ with a slowly decaying PACF suggests a moving-average MA($q$) process (both introduced in Section~\ref{sec:arma-models}) \cite{Box1970, Hyndman2021}. These plots remain indispensable even in the era of machine learning, as they provide immediate visual insight into the correlation structure of the data.

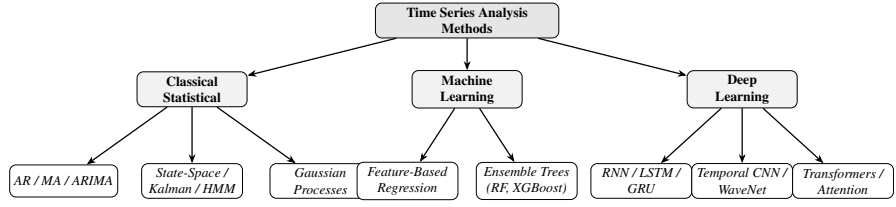
\begin{figure}[t]
\centering
\resizebox{\textwidth}{!}{%
\begin{tikzpicture}[
    every node/.style={font=\small},
    box/.style={draw, rounded corners, minimum width=2.4cm, minimum height=0.7cm, align=center, fill=white},
    root/.style={box, fill=gray!20, font=\small\bfseries, minimum width=4cm},
    cat/.style={box, fill=gray!10, font=\small\bfseries},
    leaf/.style={box, font=\small\itshape, minimum width=2.2cm},
    arr/.style={-{Stealth[length=2mm]}, thick}
]
\node[root] at (6,0) (ts) {Time Series Analysis\\Methods};
\node[cat] at (0,-1.5) (cl) {Classical\\Statistical};
\node[cat] at (6,-1.5) (ml) {Machine\\Learning};
\node[cat] at (12,-1.5) (dl) {Deep\\Learning};
\draw[arr] (ts) -- (cl);
\draw[arr] (ts) -- (ml);
\draw[arr] (ts) -- (dl);
\node[leaf] at (-2.8,-3.5) (ar) {AR / MA / ARIMA};
\node[leaf] at (0,-3.5) (ss) {State-Space /\\Kalman / HMM};
\node[leaf] at (2.8,-3.5) (gp) {Gaussian\\Processes};
\draw[arr] (cl) -- (ar);
\draw[arr] (cl) -- (ss);
\draw[arr] (cl) -- (gp);
\node[leaf] at (4.7,-3.5) (fb) {Feature-Based\\Regression};
\node[leaf] at (7.3,-3.5) (ens) {Ensemble Trees\\(RF, XGBoost)};
\draw[arr] (ml) -- (fb);
\draw[arr] (ml) -- (ens);
\node[leaf] at (9.8,-3.5) (rnn) {RNN / LSTM /\\GRU};
\node[leaf] at (12,-3.5) (cnn) {Temporal CNN /\\WaveNet};
\node[leaf] at (14.2,-3.5) (tr) {Transformers /\\Attention};
\draw[arr] (dl) -- (rnn);
\draw[arr] (dl) -- (cnn);
\draw[arr] (dl) -- (tr);
\end{tikzpicture}%
}
\caption{A taxonomy of the main approaches to time series analysis covered in this chapter, organized into three families: classical statistical methods, traditional machine learning, and deep learning architectures.}
\label{fig:taxonomy}
\end{figure}

The goals of time series analysis can be broadly categorized into several tasks. \textit{Forecasting} (prediction) is often the primary task – projecting future values of the series from its history. Other common tasks include \textit{classification} of entire series into categories (e.g. classifying a star’s variability type from its light curve), and \textit{anomaly detection} – identifying unusual or outlier sequences in time series data (e.g. a sudden burst in sensor readings or an unexpected astrophysical transient). Time series can also be used for \textit{regression} (predicting a continuous outcome from a sequence, sometimes called extrinsic regression) or \textit{clustering} (grouping similar time series). Recent surveys highlight these major applications, noting that modern approaches tackle forecasting, anomaly detection, and classification with specialized methods \cite{Hall2025}. In the following sections, we first review classical time series models developed in statistics, then move to machine learning approaches, and finally discuss advanced deep learning architectures that have gained popularity for time series analysis. Figure~\ref{fig:taxonomy} provides a visual overview of the method families covered in this chapter.

\section{Classical Time Series Models}
Before the ML era, statisticians developed a rich toolkit for modeling time series. Many of these classical methods remain highly relevant and often serve as benchmarks or components within ML pipelines.

\subsection{Autoregressive and Moving Average Models}\label{sec:arma-models}
A fundamental class of models are the linear autoregressive moving average models, epitomized by the ARIMA (Autoregressive Integrated Moving Average) family \cite{Box1970}. The key intuition is simple: the future of a time series can often be predicted as a linear combination of its recent past values and recent forecast errors. In an \textit{autoregressive} (AR) model, the value at time $t$ is regressed on its own previous values. Formally, an AR($p$) model of order $p$ is
\begin{equation}
y_t = c + \phi_1 y_{t-1} + \phi_2 y_{t-2} + \dots + \phi_p y_{t-p} + \varepsilon_t \,,
\label{eq:ar}
\end{equation}
where $\varepsilon_t$ is a noise term (white noise) \cite{Hyndman2021}. This equation says the series is explained by a linear combination of its $p$ most recent past values plus some randomness. AR models can produce a wide range of behaviors (e.g. decaying oscillations) depending on the coefficients $\phi_i$, but they assume stationarity (constraints on $\phi_i$ ensure the series does not diverge).

A \textit{moving average} (MA) model is another linear model where $y_t$ is a linear combination of past noise terms:
\begin{equation}
y_t = \mu + \theta_1 \varepsilon_{t-1} + \dots + \theta_q \varepsilon_{t-q} + \varepsilon_t \,.
\label{eq:ma}
\end{equation}
Here the current value is expressed as a sum of random shocks (errors) from the recent past. MA models capture short-term correlations (since a shock affects a few future points).

An ARMA($p,q$) model simply combines AR($p$) and MA($q$) components. However, real series often have trends or require differencing to become stationary. The ARIMA model generalizes ARMA by including an ``integration'' order $d$, which indicates how many differences are taken: in an ARIMA($p,d,q$) process the $d$-th difference of the series is modelled as an ARMA($p,q$) process. Differencing (subtracting a previous value) is a common trick to remove trends or unit roots and achieve stationarity \cite{Box1970}. For example, an ARIMA(1,1,1) model on $y_t$ means that $\Delta y_t = y_t - y_{t-1}$ (the first-differenced series) is modeled by an ARMA(1,1). Seasonal patterns can be handled with seasonal ARIMA (SARIMA), which includes seasonal differencing and seasonal AR/MA terms for periodic effects \cite{Hamilton1994}.

When multiple time series are observed simultaneously and are expected to influence each other, the natural extension of the AR model is the \textit{Vector Autoregression} (VAR) model \cite{Hamilton1994}. In a VAR($p$) model, each variable at time $t$ is a linear function of the past $p$ values of \textit{all} variables in the system:
\begin{equation}
\mathbf{y}_t = \mathbf{c} + \mathbf{A}_1 \mathbf{y}_{t-1} + \dots + \mathbf{A}_p \mathbf{y}_{t-p} + \boldsymbol{\varepsilon}_t \,,
\label{eq:var}
\end{equation}
where $\mathbf{y}_t$ is a vector of observations and $\mathbf{A}_i$ are coefficient matrices. VAR models are the workhorse of multivariate time series analysis in econometrics and have applications wherever cross-correlations between series are important. In astrophysics, a VAR-like framework could model correlated variability across multiple wavelength bands of a single source, or the interplay between different cosmological observables evolving in time.

Despite being linear, ARIMA models have been extremely popular for forecasting across industries due to their simplicity and solid theoretical foundation. They can often provide surprisingly strong forecasts for many practical problems, especially those with strong autocorrelations and seasonal patterns. For example, ARIMA and its seasonal variants are widely used in economics and finance for market forecasting, in supply chain management for demand planning, and even in climatology for weather variable prediction. ARIMA models have been applied to short-term weather forecasts (temperature, precipitation), to economic indicators (modeling GDP or inflation), and to hospital patient admissions – essentially any context where an extrapolation of a time-dependent trend is needed. In astronomy, while ARIMA is less commonly used (since many astrophysical processes are not easily approximated as linear models), one could imagine using ARIMA as a baseline to predict the next data point in a smoothly varying light curve.

Another cornerstone of classical methods is \textit{exponential smoothing}. Exponential smoothing forecasts a series by weighting past observations with exponentially decaying weights, so that more recent observations have greater influence. Simple exponential smoothing is effective for series without clear trend or seasonality; Holt’s linear method adds a trend component, and Holt-Winters method adds seasonal components. These methods have been very successful in business forecasting (for inventory and sales) due to their robustness and ease of use. For example, the Holt-Winters method (also known as triple exponential smoothing) is a popular choice for forecasting seasonal data like retail sales and electricity demand, and was a top performer in some forecasting competitions prior to the prevalence of machine learning methods \cite{Hyndman2021}. Unlike ARIMA, exponential smoothing is more heuristic and was historically separate from the ARIMA family, but interestingly it was shown that many exponential smoothing methods can be derived from state-space models and have ARIMA equivalents. For practitioners, \textit{Prophet} \cite{Taylor2018} offers an accessible decomposition-based interface combining trend, seasonality, and holiday effects.

\subsection{Spectral Methods and Period Finding}
An alternative to modeling a time series in the time domain is to analyze it in the \textit{frequency domain}. Spectral analysis decomposes a signal into its constituent frequencies, revealing periodic components that may not be immediately obvious in the raw data. The standard tool is the \textit{power spectral density} (PSD), estimated via the periodogram or its smoothed variants. A peak in the PSD at a given frequency indicates a periodic component at that period.

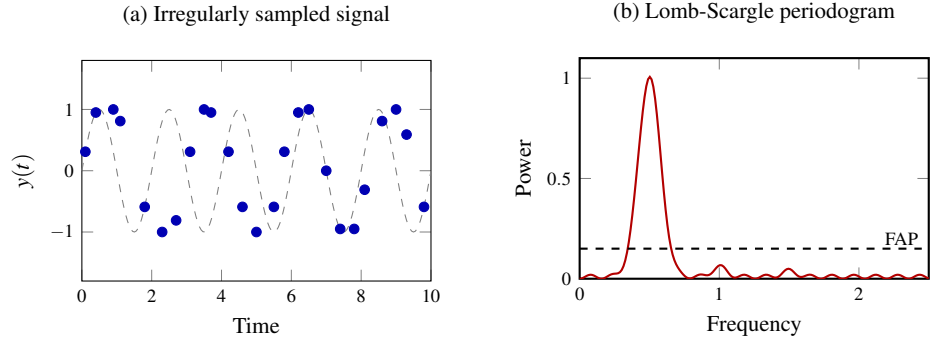
\begin{figure}[t]
\centering
\begin{tikzpicture}
\begin{axis}[
    name=left,
    width=6.2cm, height=4.5cm,
    xlabel={Time}, ylabel={$y(t)$},
    xlabel style={font=\small}, ylabel style={font=\small},
    tick label style={font=\scriptsize},
    title={\small (a) Irregularly sampled signal},
    title style={at={(0.5,1.05)}},
    xmin=0, xmax=10, ymin=-1.8, ymax=1.8
]
\addplot[gray, dashed, no markers, thin, domain=0:10, samples=200]
    {sin(deg(2*3.14159*0.5*x))};
\addplot[only marks, mark=*, mark size=1.8pt,
         color=blue!70!black, mark options={fill=blue!70!black, draw=blue!70!black}]
    coordinates {
    (0.1,0.31) (0.4,0.95) (0.9,1.00) (1.1,0.81)
    (1.8,-0.59) (2.3,-1.00) (2.7,-0.81) (3.1,0.31)
    (3.5,1.00) (3.7,0.95) (4.2,0.31) (4.6,-0.59)
    (5.0,-1.00) (5.5,-0.59) (5.8,0.31) (6.2,0.95)
    (6.5,1.00) (7.0,0.00) (7.4,-0.95) (7.8,-0.95)
    (8.1,-0.31) (8.6,0.81) (9.0,1.00) (9.3,0.59)
    (9.8,-0.59)
};
\end{axis}
\begin{axis}[
    name=right, at={(left.outer east)}, anchor=outer west,
    xshift=0.8cm,
    width=6.2cm, height=4.5cm,
    xlabel={Frequency}, ylabel={Power},
    xlabel style={font=\small}, ylabel style={font=\small},
    tick label style={font=\scriptsize},
    title={\small (b) Lomb-Scargle periodogram},
    title style={at={(0.5,1.05)}},
    xmin=0, xmax=2.5, ymin=0, ymax=1.1,
    no markers, thick
]
\addplot[red!70!black, domain=0:2.5, samples=200]
    {exp(-80*(x-0.5)^2) + 0.05*exp(-200*(x-1.0)^2) + 0.03*exp(-200*(x-1.5)^2) + 0.02*sin(deg(20*x))^2};
\draw[dashed, thick, black] (axis cs:0,0.15) -- (axis cs:2.45,0.15)
    node[above left, font=\scriptsize, inner sep=1pt] {FAP};
\end{axis}
\end{tikzpicture}
\caption{Illustration of Lomb-Scargle period analysis. (a) A sinusoidal signal with frequency $f=0.5$ (dashed gray curve) sampled at irregular time intervals (blue dots), as is typical of astronomical observations. (b) The resulting Lomb-Scargle periodogram shows a clear power peak at the true frequency, well above the false-alarm probability (FAP) threshold (dashed line).}
\label{fig:lombscargle}
\end{figure}

For regularly sampled data, the classical Fourier-based periodogram is straightforward. However, most astronomical time series are sampled irregularly due to observing constraints (weather, day/night cycles, telescope scheduling). For such data, the \textit{Lomb-Scargle periodogram} \cite{Lomb1976, Scargle1982, VanderPlas2018} is the standard tool. It generalizes the classical periodogram to unevenly spaced data by fitting sinusoids at each trial frequency. For a time series $\{y(t_i)\}$ with mean $\bar{y}$, the Lomb-Scargle power at angular frequency $\omega$ is:
\begin{equation}
P(\omega) = \frac{1}{2} \left[ \frac{\left(\sum_i (y_i - \bar{y}) \cos\omega(t_i - \tau)\right)^2}{\sum_i \cos^2\omega(t_i - \tau)} + \frac{\left(\sum_i (y_i - \bar{y}) \sin\omega(t_i - \tau)\right)^2}{\sum_i \sin^2\omega(t_i - \tau)} \right] \,,
\label{eq:lombscargle}
\end{equation}
where $\tau$ is a time offset chosen to make the cosine and sine terms orthogonal. Figure~\ref{fig:lombscargle} illustrates the method on a sinusoidal signal sampled at irregular times. The Lomb-Scargle periodogram is ubiquitous in astronomy: it is routinely used to determine the periods of variable stars, detect orbital signals of exoplanets in radial velocity data, and identify quasi-periodic oscillations in X-ray binaries and active galactic nuclei. Its statistical properties are well understood, allowing for rigorous false-alarm probability assessments~\cite{Scargle1982}. Modern implementations (e.g.\ in \texttt{astropy}\footnote{\texttt{astropy} is a community-developed Python library for astronomy: \url{https://www.astropy.org} \cite{Astropy2022}.}) are fast and widely accessible.

Spectral methods and time-domain models are complementary: spectral analysis excels at identifying periodicities and characterizing the frequency content of a signal, while ARIMA-type models are better suited for forecasting and capturing autoregressive dynamics. In practice, a period found via Lomb-Scargle can inform the seasonal component of a SARIMA model or serve as a feature for a machine learning classifier.

Both Fourier and Lomb-Scargle analyses assume stationary frequency content. For transient and non-stationary signals -- gravitational-wave chirps, stellar flares, or AGN variability spanning many timescales -- a \emph{time--frequency} decomposition is more informative. The continuous wavelet transform (CWT), typically using a Morlet wavelet, convolves the series with scaled and shifted copies of a localised template, producing a 2-D \emph{scalogram} of power in $(t,\,\text{scale})$ \cite{Torrence1998}. Scalograms and spectrograms are themselves image-like objects, which is why they are natural inputs for the 2-D Convolutional Neural Networks (CNNs) discussed in Section~\ref{sec:cnn}.

\subsection{State-Space and Probabilistic Models}
State-space models provide another perspective by introducing latent (hidden) state variables that evolve over time. The most famous example is the \textit{Kalman filter}, which is essentially a state-space formulation of a linear dynamical system with Gaussian noise \cite{Kalman1960}. In a Kalman filter model, one assumes there is an unobserved state $\mathbf{z}_t$ that follows a linear evolution, and the observations are a linear function of the state:
\begin{align}
\mathbf{z}_t &= \mathbf{A}\, \mathbf{z}_{t-1} + \boldsymbol{\omega}_t \,, & \boldsymbol{\omega}_t &\sim \mathcal{N}(\mathbf{0}, \mathbf{Q}) \,, \label{eq:kalman_state}\\
\mathbf{y}_t &= \mathbf{C}\, \mathbf{z}_t + \boldsymbol{\nu}_t \,, & \boldsymbol{\nu}_t &\sim \mathcal{N}(\mathbf{0}, \mathbf{R}) \,, \label{eq:kalman_obs}
\end{align}
where $\mathbf{Q}$ and $\mathbf{R}$ are the process and observation noise covariance matrices, respectively. The Kalman filter provides an efficient recursive algorithm to estimate the state from data and predict future observations. Kalman filters (and their extensions) are widely used in engineering (e.g. tracking objects, sensor fusion in robotics) and also in astrophysics for tasks like orbit determination or spacecraft navigation. They are optimal for linear Gaussian systems but can be extended to non-linear cases (Extended Kalman Filter, Unscented KF) and are a foundation for many modern sequential Bayesian filtering techniques.

Another early machine-learning approach (before the term "ML" was common) for time series is the \textit{Hidden Markov Model} (HMM) \cite{Rabiner1989}. HMMs were introduced in the late 1960s and became very popular for speech recognition in the 1980s, modeling speech signals as probabilistic transitions between a set of discrete hidden states. An HMM assumes there is a hidden state process that is a Markov chain (each state depends only on the previous state), and observations are generated probabilistically from each state. The classic use-case is to model sequences where there are distinct regimes or patterns. For example, in speech, the hidden state might correspond to phonemes; in finance, one might use an HMM to model market regimes (bull vs bear markets) that switch probabilistically. In astrophysics, HMMs have been used occasionally, for instance to model mode switching in pulsar emissions or characterize stages of variability in X-ray binary light curves. HMMs can be trained using the Baum-Welch algorithm (an Expectation-Maximization method) and decoded with the Viterbi algorithm to find the most likely state sequence. They remain useful for certain problems, though have been largely supplanted by more flexible, non-linear models -- notably recurrent and deep neural networks, discussed in Section~\ref{sec:dl} -- in many applications.

A powerful probabilistic approach for time series regression is the \textit{Gaussian Process} (GP). Gaussian processes provide a flexible, non-parametric way to define a distribution over functions, and they have become a favored tool in fields where quantifying uncertainty is important, including scientific domains like astronomy. Formally, a GP is fully specified by a mean function $m(t) = \mathbb{E}[f(t)]$ and a covariance (kernel) function $k(t, t') = \text{Cov}(f(t), f(t'))$. Given observed data $\mathbf{y}$ at times $\mathbf{t}$, the predictive distribution at a new time $t_*$ is Gaussian with mean and variance:
\begin{align}
\mu_* &= \mathbf{k}_*^\top (\mathbf{K} + \sigma_n^2 \mathbf{I})^{-1} \mathbf{y} \,, \label{eq:gp_mean}\\
\sigma_*^2 &= k(t_*, t_*) - \mathbf{k}_*^\top (\mathbf{K} + \sigma_n^2 \mathbf{I})^{-1} \mathbf{k}_* \,, \label{eq:gp_var}
\end{align}
where $\mathbf{K}$ is the kernel matrix with entries $K_{ij} = k(t_i, t_j)$, $\mathbf{k}_*$ is the vector of covariances between $t_*$ and the observed times, and $\sigma_n^2$ is the noise variance \cite{Rasmussen2006}. By choosing a kernel function that encodes assumptions about the data (e.g.\ smoothness, periodicity), one can fit complex functions to data and obtain a predictive mean and confidence interval for future points. GPs can be seen as a Bayesian approach to curve fitting. In astronomy, GPs have recently emerged as a method of choice for modeling stochastic variability in time-domain data due to their flexibility and principled uncertainty estimates \cite{Aigrain2023}. For example, GPs are used in exoplanet research to model stellar activity (which is a noisy time series that can obscure planet signals) and in gravitational lensing or quasar light curves to model irregular variations. Aigrain \& Foreman-Mackey (2023) review numerous such applications, noting that GPs combine mathematical simplicity, flexibility in modeling complex temporal covariance, and robustness, making them ideal for many astronomical time series problems \cite{Aigrain2023}. The downside is computational: naive GP inference scales poorly with the number of data points, which can be a challenge for very large datasets (astronomical light curves can have $10^{5}$--$10^{6}$ observations per object in LSST-era surveys \cite{Ivezic2019}). Nonetheless, with sparse GP approximations and faster software, GPs are increasingly feasible even as data grow.

\subsection{Continuous-Time Autoregressive Models (CARMA)}\label{sec:carma}
All the autoregressive models introduced so far assume observations on a regular time grid with fixed spacing $\Delta t$. In astronomy this assumption is routinely violated: ground-based light curves have gaps from weather, scheduling, and the day/night cycle, while space-based missions leave gaps due to orbital constraints or telemetry dropouts. Discretising an irregularly sampled series by binning or interpolation discards information and introduces biases, so a natural alternative is to model the series directly in continuous time. The \emph{Continuous-time AutoRegressive Moving Average} (CARMA) family \cite{Kelly2014} does exactly this, reformulating AR/MA as a stochastic differential equation (SDE). A CARMA($p,q$) process satisfies
\begin{equation}
\frac{d^{p} y}{dt^{p}} + \alpha_{p-1}\frac{d^{p-1} y}{dt^{p-1}} + \cdots + \alpha_{0}\, y(t)
\;=\; \beta_{q}\frac{d^{q}\varepsilon}{dt^{q}} + \cdots + \beta_{0}\,\varepsilon(t)\,,
\label{eq:carma}
\end{equation}
where $\varepsilon(t)$ is a continuous-time white-noise driving process. The CARMA likelihood at arbitrary (non-uniform) observation times admits a closed form, so no resampling of the data is needed.

The simplest non-trivial instance, CARMA(1,0), is the \emph{damped random walk} (DRW), equivalent to an Ornstein--Uhlenbeck process:
\begin{equation}
dy = -\frac{1}{\tau}\left(y - \mu\right)dt + \sigma\, dW\,,
\label{eq:drw}
\end{equation}
with characteristic decorrelation timescale $\tau$ and long-term variance $\sigma^{2}\tau/2$. The DRW has an exponential autocorrelation $\exp(-|t|/\tau)$ and a Lorentzian power spectrum, and it has become the standard phenomenological model for the optical variability of quasars and active galactic nuclei \cite{Kelly2009, MacLeod2010}. Higher-order members of the family are also physically meaningful: CARMA(2,1), for example, describes a damped stochastic harmonic oscillator and is used to capture quasi-periodic variability in X-ray binaries and AGN \cite{Kelly2014}.

What ties CARMA back to Section~\ref{sec:arma-models} and to the state-space / Gaussian-process material just discussed is that every CARMA model admits a \emph{linear state-space representation}. The Kalman filter of Eqs.~\eqref{eq:kalman_state}--\eqref{eq:kalman_obs} then evaluates the likelihood in $\mathcal{O}(N)$ time (where $N$ is the number of observations), in contrast to the $\mathcal{O}(N^{3})$ cost of a naive GP fit to Eqs.~\eqref{eq:gp_mean}--\eqref{eq:gp_var}. This equivalence is exploited by the \texttt{celerite} and \texttt{celerite2} libraries \cite{ForemanMackey2017}, whose kernels are built from CARMA(2,1)-style components; it is what makes principled Bayesian modelling of astronomical light curves feasible at the $10^{5}$--$10^{6}$ observations per object expected in the LSST era. The main caveats are that CARMA assumes Gaussian driving noise and (weak) stationarity, so intrinsically non-stationary events -- stellar flares, supernova explosions, AGN state transitions -- typically require either piecewise modelling or non-stationary extensions of the framework.

In summary, classical approaches like ARIMA, exponential smoothing, state-space models (Kalman filters), HMMs, GPs, and continuous-time CARMA/DRW models each bring different strengths. ARIMA and exponential smoothing are fast, interpretable, and often quite effective for short-term forecasting in stationary or seasonal contexts. State-space models and HMMs provide a dynamical-systems view and are useful when an underlying process with states or memory is suspected. Gaussian processes offer a flexible Bayesian framework to capture complex correlations with uncertainty quantification, and CARMA/DRW models extend this picture to continuous time, natively handling irregularly sampled series at $\mathcal{O}(N)$ cost. However, these approaches also have limitations: many (like ARIMA) are inherently linear and may struggle with nonlinear patterns; others (like HMMs) might oversimplify by using discrete states; and some (GPs) become computationally intensive with large data. These limitations opened the door for more powerful machine learning methods to handle time series, which are discussed next.

\section{Machine Learning Approaches for Time Series}
Machine learning methods for time series aim to overcome some limitations of classical models by either automating the feature extraction or by using flexible nonlinear models that can capture complex patterns in data. We can loosely divide ML approaches into two groups: (1) techniques that transform the time series into features for use with generic ML algorithms (feature-based approaches), and (2) specialized ML models that directly handle sequential data (including deep learning models, covered in the next section).

\subsection{Feature-Based and Ensemble Methods}\label{sec:feature-based}
One straightforward approach to apply machine learning is to convert the time series prediction problem into a supervised learning problem by extracting features from past observations. For example, to forecast a value $y_{t+h}$ (h steps ahead), one can take the recent history $(y_t, y_{t-1}, \dots, y_{t-k})$ and various summary statistics as input features to a regression model. These features might include lags (the values $y_{t-1}, y_{t-2}, \dots$ themselves), moving averages, recent trend estimates, seasonal indicators (to encode the time of year or day), etc. Once the time series is represented in a feature vector form, any regression algorithm can be used: linear regression, support vector machines, decision trees, random forests, gradient boosting machines, etc  \cite{Breiman2001}. This approach does not explicitly use a sequential model, but rather relies on the feature engineering to capture temporal patterns.

Ensemble tree-based methods, such as \textit{Random Forests} and \textit{Gradient Boosting Machines} (e.g. XGBoost, LightGBM, CatBoost) \cite{Chen2016, Ke2017, Prokhorenkova2018}, have proven extremely effective in many time series forecasting competitions and applications. A recent survey of top-performing time series prediction models found that gradient-boosted decision tree algorithms (like LightGBM) often achieve excellent accuracy across diverse forecasting tasks, on par with or better than many deep learning models, while also being more computationally efficient in training \cite{Hall2025}. These tree-based models can naturally handle nonlinear relationships and interactions between features, and they are fairly robust to outliers and missing data – all useful properties for real-world time series. For example, in the M5 forecasting competition \cite{Makridakis2022M5} (which involved predicting retail sales for thousands of products), many of the winning solutions used carefully engineered features fed into XGBoost/LightGBM models. Tree models can easily incorporate external (exogenous) features as well, such as holiday flags or weather data for energy demand forecasting. Indeed, an ARIMAX model in classical terms is analogous to including exogenous inputs, whereas in ML one simply adds more features and lets the model figure out the relationships. One should note, however, that these models do not inherently account for the temporal order beyond what is coded in the features, so one must be careful to include enough lag features and relevant context.

\begin{figure}[t]
\centering
\begin{tikzpicture}[
    font=\small,
    train/.style={fill=blue!30, draw=blue!50!black, minimum height=0.45cm},
    test/.style={fill=red!30, draw=red!50!black, minimum height=0.45cm},
    unused/.style={fill=gray!10, draw=gray!50, minimum height=0.45cm}
]
\node[anchor=west, font=\small\bfseries] at (-0.5, 2.8) {(a) Standard $k$-fold CV (incorrect for time series)};
\foreach \fold/\tst in {1/1, 2/3, 3/5} {
    \pgfmathsetmacro{\ypos}{2.4 - (\fold-1)*0.6}
    \node[anchor=east, font=\scriptsize] at (-0.1, \ypos) {Fold \fold};
    \foreach \blk in {1,...,6} {
        \pgfmathsetmacro{\xpos}{(\blk-1)*1.5}
        \ifnum\blk=\tst
            \node[test, minimum width=1.4cm] at (\xpos+0.7, \ypos) {};
        \else
            \node[train, minimum width=1.4cm] at (\xpos+0.7, \ypos) {};
        \fi
    }
}
\draw[-{Stealth[length=2mm]}, thick] (0, 0.55) -- (9.0, 0.55) node[right, font=\scriptsize] {Time};

\node[anchor=west, font=\small\bfseries] at (-0.5, -0.2) {(b) Walk-forward CV (correct)};
\foreach \fold/\trainend in {1/3, 2/4, 3/5} {
    \pgfmathsetmacro{\ypos}{-0.6 - (\fold-1)*0.6}
    \pgfmathsetmacro{\testblk}{\trainend+1}
    \node[anchor=east, font=\scriptsize] at (-0.1, \ypos) {Fold \fold};
    \foreach \blk in {1,...,6} {
        \pgfmathsetmacro{\xpos}{(\blk-1)*1.5}
        \pgfmathtruncatemacro{\te}{\testblk}
        \ifnum\blk=\te
            \node[test, minimum width=1.4cm] at (\xpos+0.7, \ypos) {};
        \else
            \ifnum\blk>\te
                \node[unused, minimum width=1.4cm] at (\xpos+0.7, \ypos) {};
            \else
                \node[train, minimum width=1.4cm] at (\xpos+0.7, \ypos) {};
            \fi
        \fi
    }
}
\draw[-{Stealth[length=2mm]}, thick] (0, -2.45) -- (9.0, -2.45) node[right, font=\scriptsize] {Time};
\node[train, minimum width=0.8cm] at (2.5, -3.0) {};
\node[anchor=west, font=\scriptsize] at (3.0, -3.0) {Train};
\node[test, minimum width=0.8cm] at (4.8, -3.0) {};
\node[anchor=west, font=\scriptsize] at (5.3, -3.0) {Test};
\node[unused, minimum width=0.8cm] at (7.0, -3.0) {};
\node[anchor=west, font=\scriptsize] at (7.5, -3.0) {Unused};
\end{tikzpicture}
\caption{Comparison of cross-validation strategies for time series. (a) Standard $k$-fold CV randomly assigns time blocks to train and test sets, allowing the model to ``see the future'' (data leakage). (b) Walk-forward (expanding window) CV always trains on past data and tests on the next block, preserving causal ordering.}
\label{fig:cv}
\end{figure}
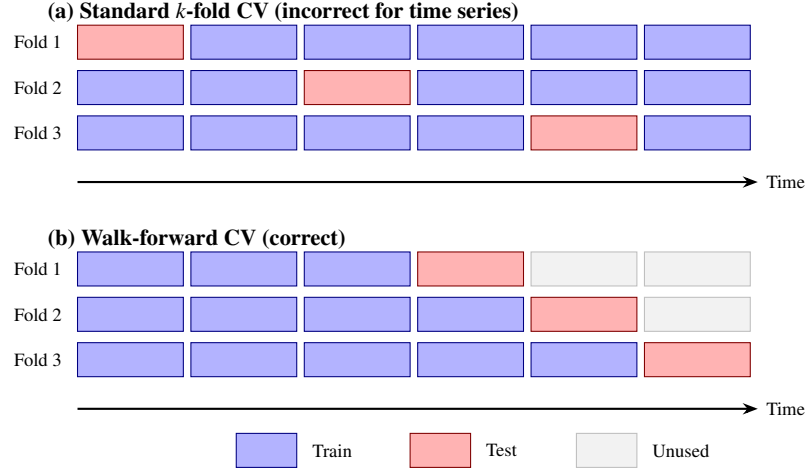

A critical practical consideration when applying any ML model to time series is \textit{evaluation methodology}. Standard $k$-fold cross-validation, which randomly shuffles data into training and test folds, is inappropriate for time series because it breaks the temporal ordering and introduces \textit{data leakage}: the model may train on future data to predict the past, producing overly optimistic performance estimates. The correct approach is \textit{time series cross-validation} (also called walk-forward validation or expanding/sliding window validation). In this scheme, the training set always consists of observations \textit{before} the test set in time. For example, one trains on months 1--12, tests on month 13; then trains on months 1--13, tests on month 14; and so on (see Fig.~\ref{fig:cv}). This respects the causal structure of the data and produces realistic estimates of out-of-sample performance \cite{Hyndman2021}. Failure to use proper temporal validation is one of the most common pitfalls when applying machine learning to time series in practice.

A complementary strategy is \textit{automated feature extraction}. Libraries such as \texttt{tsfresh} \cite{Christ2018} systematically compute hundreds of statistical features from a time series (e.g.\ entropy, number of peaks, autoregressive coefficients, wavelet energies) and then use relevance filtering to select the most informative ones. This approach makes it possible to apply standard classifiers or regressors to time series data with minimal manual feature engineering, and has been adopted in fields ranging from predictive maintenance to medical diagnostics.

Another traditional approach related to features is using \textit{distance-based} or \textit{instance-based} learning for time series. For classification tasks, a classic method is the nearest-neighbor classifier under a time-series similarity measure like Dynamic Time Warping (DTW) \cite{Berndt1994}. DTW measures similarity between time series that may be stretched or misaligned in time by optimally warping the time axis. A 1-NN classifier with DTW distance was a strong baseline for time series classification for many years. Some approaches learn shape-based features (so-called \textit{shapelets}) \cite{Ye2009} that are small subsequences particularly discriminative of classes.

More recently, the \textit{ROCKET} (RandOm Convolutional KErnel Transform) family of methods \cite{Dempster2020} has emerged as a state-of-the-art approach for time series classification. ROCKET applies a large number of random convolutional kernels to the input series, extracts simple summary statistics (max value and proportion of positive values) from each convolution output, and feeds the resulting feature vector into a linear classifier. Despite its simplicity, ROCKET achieves accuracy comparable to deep learning models at a fraction of the computational cost. Its successor, MiniRocket, further reduces computation while maintaining accuracy, making it practical for very large datasets. In astronomy, these methods are promising for the rapid classification of transients in survey pipelines where computational speed is critical.

In astronomy, one could also classify variable star light curves by comparing them to templates or prototypical patterns (using DTW to account for slight differences in period or phase). While effective in some cases, nearest-neighbor methods can be slow for large databases, though indexing methods and lower bounds for DTW exist to speed up retrieval.

\subsection{Hybrid and Specialized Models}
Many real-world applications benefit from combining the strengths of classical and ML approaches. For example, one can use an ARIMA model to handle seasonal patterns and trend, and feed its residuals (which ideally contain only more complex variations) into a machine learning model to capture any remaining nonlinear structure. This is sometimes called ARIMA residual learning or hybrid modeling. In a different vein, another hybrid approach is to use machine learning to choose among or combine forecasts from multiple classical models (sometimes via stacking or meta-learning).

Another area of development has been in tailored ML models for time series. For example, \textit{shapelet transformation} methods explicitly search for small subsequences that best differentiate classes, and then use those distances as features in a classifier. Methods like \textit{time series forest} create random interval features from the series (e.g., mean, standard deviation over random time intervals) and build an ensemble of trees, which has proven to be a competitive approach for classification. These methods often blend the line between manual feature engineering and automated feature learning.

It is also worth mentioning the increasing role of \textit{unsupervised learning} on time series. Techniques such as clustering or anomaly detection often proceed by defining a distance measure between series (like DTW or correlation-based distances) and then using algorithms like k-means or DBSCAN to cluster similar time series or to identify outliers. Anomaly detection can also be tackled by one-class classification methods or by building forecasting models and flagging large prediction residuals as anomalies. In astronomy, anomaly detection is crucial for finding novel transient events or unusual variable stars in surveys; ML approaches like autoencoders (discussed later) and clustering in feature space have been used to let the data itself indicate what's "normal" versus "odd" without having to manually label anomalies.

\subsection{Evaluation Metrics}\label{sec:evaluation-metrics}
Choosing appropriate evaluation metrics is essential for meaningful model comparison. For point forecasting, the most common metrics are the Mean Absolute Error (MAE $= \frac{1}{n}\sum_{i=1}^{n} |y_i - \hat{y}_i|$), Root Mean Squared Error (RMSE $= \sqrt{\frac{1}{n}\sum_{i=1}^{n} (y_i - \hat{y}_i)^2}$), and Mean Absolute Percentage Error (MAPE $= \frac{100}{n}\sum_{i=1}^{n} |y_i - \hat{y}_i|/|y_i|$), where $n$ is the number of test observations and $\hat{y}_i$ the forecast for the $i$-th observation. MAPE is scale-independent and widely used in business forecasting, but is undefined when $y_i = 0$ and can be asymmetric; the symmetric MAPE (sMAPE) and the Mean Absolute Scaled Error (MASE) \cite{Hyndman2006} address some of these limitations. MASE, in particular, scales errors relative to a naive forecast and is recommended for comparing accuracy across series with different scales.

For probabilistic forecasts, which provide full predictive distributions rather than point estimates, proper scoring rules such as the \textit{Continuous Ranked Probability Score} (CRPS) are preferred. CRPS generalizes MAE to probabilistic forecasts and rewards both calibration (are the predicted intervals reliable?) and sharpness (are they narrow?). In scientific applications where uncertainty quantification is critical -- such as predicting the time of next eruption of a variable star or constraining cosmological parameters from time series data -- probabilistic metrics should be used alongside point metrics.

For classification tasks, standard metrics apply: accuracy, precision, recall, $F_1$-score, and the area under the ROC curve (AUC-ROC). In astronomy, where class imbalance is common (e.g., rare transients among millions of normal sources), precision-recall curves and the AUC-PR are often more informative than accuracy alone.

Overall, classical ML approaches (feature-based, distance-based, ensembles) have enriched the time series toolkit by enabling flexible nonlinear modeling and data-driven feature extraction. However, they often still rely on human expertise to choose the right features or transformations of the time axis. The next major leap in time series analysis came with the rise of deep learning, which aims to automatically learn the relevant features and complex patterns directly from raw sequential data.

\section{Deep Learning for Time Series}\label{sec:dl}
In recent years, deep learning has revolutionized many domains by enabling end-to-end learning of complex patterns from raw data. Time series are no exception. Deep neural networks, with their ability to approximate highly nonlinear functions, have been applied to time series tasks with considerable success. The most common deep learning architectures for sequential data are Recurrent Neural Networks (RNNs), Convolutional Neural Networks (CNNs), and more recently Transformer networks with attention mechanisms. Each offers different advantages for modeling time series, and often they are used in combination.

\subsection{Recurrent Neural Networks and LSTMs}\label{sec:rnn}
Recurrent neural networks are specifically designed to handle sequential inputs by maintaining an internal state (memory) that is updated at each time step. Unlike a standard feed-forward network that assumes inputs are independent, an RNN processes data one step at a time, using the output (hidden state) from the previous step as an additional input to the next. This creates a chain-like dependency that allows the network to retain information from the past. Conceptually, the RNN maintains a ``memory'' through a feedback loop: its output at each step is fed back as part of the input at the next step. This recurrent structure makes RNNs a natural architecture for time series data and sequential data more broadly.

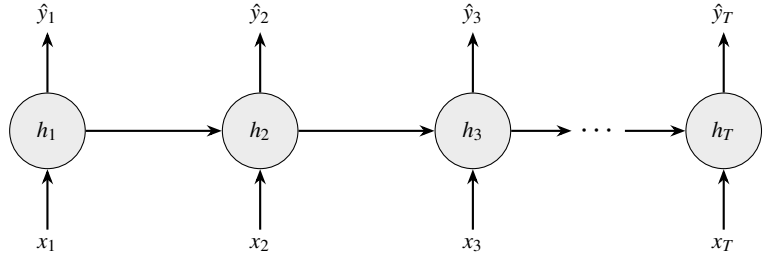
\begin{figure}[t]
\centering
\begin{tikzpicture}[
    node distance=1.8cm and 1.8cm,
    every node/.style={font=\small},
    cell/.style={draw, circle, minimum size=1.0cm, fill=gray!15, font=\small\bfseries},
    io/.style={font=\small},
    arr/.style={-{Stealth[length=2mm]}, thick}
]
\node[cell] (h1) {$h_1$};
\node[cell, right=of h1] (h2) {$h_2$};
\node[cell, right=of h2] (h3) {$h_3$};
\node[right=0.8cm of h3, font=\large] (dots) {$\cdots$};
\node[cell, right=0.8cm of dots] (hT) {$h_T$};

\node[io, below=0.8cm of h1] (x1) {$x_1$};
\node[io, below=0.8cm of h2] (x2) {$x_2$};
\node[io, below=0.8cm of h3] (x3) {$x_3$};
\node[io, below=0.8cm of hT] (xT) {$x_T$};

\node[io, above=0.8cm of h1] (y1) {$\hat{y}_1$};
\node[io, above=0.8cm of h2] (y2) {$\hat{y}_2$};
\node[io, above=0.8cm of h3] (y3) {$\hat{y}_3$};
\node[io, above=0.8cm of hT] (yT) {$\hat{y}_T$};

\draw[arr] (x1) -- (h1);
\draw[arr] (x2) -- (h2);
\draw[arr] (x3) -- (h3);
\draw[arr] (xT) -- (hT);

\draw[arr] (h1) -- (y1);
\draw[arr] (h2) -- (y2);
\draw[arr] (h3) -- (y3);
\draw[arr] (hT) -- (yT);

\draw[arr] (h1) -- (h2);
\draw[arr] (h2) -- (h3);
\draw[arr] (h3) -- (dots);
\draw[arr] (dots) -- (hT);
\end{tikzpicture}
\caption{Schematic of a Recurrent Neural Network (RNN) unrolled in time. At each time step $t$, the network receives an input $x_t$ and the previous hidden state $h_{t-1}$, producing an updated hidden state $h_t$ and an output $\hat{y}_t$. The horizontal arrows represent the flow of information through time, enabling the network to maintain a memory of past inputs.}
\label{fig:rnn}
\end{figure}

Formally, a basic RNN can be described as follows: at each time $t$, the network takes the current input $x_t$ (which could be the value of the time series at $t$ or a vector of observations at time $t$) and the previous hidden state $h_{t-1}$, and computes a new hidden state $h_t = f(W x_t + U h_{t-1} + b)$ for some nonlinear activation function $f$ (like tanh) and weight matrices $W, U$. The hidden state $h_t$ can be thought of as an encoding of all relevant information seen up to time $t$. The network may also produce an output $y_t$ -- typically a prediction -- based on $h_t$. A schematic of an RNN unrolled in time is shown in Fig.~\ref{fig:rnn}. Through training on sequential data, the RNN learns to update its hidden state in a way that captures patterns in the sequence and to use that to make predictions.

RNNs were actually introduced decades ago, but only became truly effective with the advent of better training techniques and more data. One key development was the \textit{Long Short-Term Memory} (LSTM) network, introduced by Hochreiter \& Schmidhuber (1997) to address the notorious "vanishing gradient" problem that made training basic RNNs difficult for long sequences \cite{Hochreiter1997}. Standard RNNs tend to have trouble learning long-term dependencies – they might be good at using information from, say, the last 5 steps, but struggle if the relevant information was 50 steps ago. LSTMs mitigate this by using a special gated architecture with separate mechanisms (gates) to control what information is kept, forgotten, or output at each time step \cite{Hochreiter1997}. An LSTM unit has an internal cell state that can carry information across many time steps, and gates (input, output, forget gates) that regulate the flow of information. This design allows gradients to flow better during training, enabling the network to learn relationships over longer horizons. A variant called the \textit{Gated Recurrent Unit} (GRU) \cite{Cho2014} simplifies the LSTM architecture by using only two gates (reset and update) and has also become popular. GRUs often perform comparably to LSTMs with fewer parameters.

In practice, RNNs (especially LSTMs/GRUs) became the workhorse for many time series and sequence modeling tasks in the 2010s. For example, in forecasting, an LSTM can be trained to ingest a sequence of past values and directly output a sequence of future values (sequence-to-sequence forecasting \cite{Sutskever2014}). In anomaly detection, an LSTM-based autoencoder \cite{Malhotra2015} can be trained to reconstruct normal time series sequences and flag those with large reconstruction error as anomalies. In astronomy, RNNs have been used for classifying light curves of variable stars and transients – one can feed the sequential brightness measurements into an LSTM and have it output a class label at the end. Because RNNs do not assume linearity or stationarity and can, in principle, approximate very complex functions, they can model phenomena that classical models cannot. For instance, an LSTM could learn the irregular pattern of a star that has multiple outburst states (something a single ARIMA model would struggle with). RNNs have also been applied in cosmology, for example to emulate expensive physics simulations by learning from time series of simulation data, or to predict the evolution of cosmological parameters.

An important practical consideration for any forecasting model -- but especially relevant for deep learning -- is the distinction between \textit{single-step} and \textit{multi-step} forecasting. Given observations up to time $t$, the \textit{forecast horizon} $H$ is the number of future time steps we wish to predict. In single-step forecasting ($H=1$) the model predicts only the next time step $y_{t+1}$. For predictions further into the future ($H>1$) there are three main strategies: (i) the \textit{recursive} (or iterated) strategy, where the model's own prediction is fed back as input to generate the next, cascading errors as we move further along the horizon; (ii) the \textit{direct} strategy, where a separate model is trained for each lead time $h = 1, \ldots, H$, avoiding error accumulation but ignoring dependencies between different lead times; and (iii) the \textit{multi-output} (or sequence-to-sequence) strategy, where a single model outputs the entire forecast vector $(y_{t+1}, \ldots, y_{t+H})$ at once. The multi-output approach, naturally suited to encoder-decoder architectures, has become the dominant paradigm in deep learning for time series, as it balances computational efficiency with the ability to capture inter-horizon dependencies.

However, RNNs are not without limitations. Training them is computationally intensive and can be unstable: because the network is unrolled over many time steps, back-propagated gradients are effectively multiplied by long chains of weight matrices and can grow without bound (``exploding gradients'') or shrink to zero (``vanishing gradients''), making the loss diverge or stall. Careful tuning and gradient clipping (capping gradient norms during training) are typically needed to mitigate these effects. They also inherently process one time step after another, which makes them difficult to parallelize on hardware (unlike CNNs, for example, which can process elements of a sequence in parallel to some extent). By the late 2010s, even though LSTM/GRU networks were dominant for tasks like language modeling and speech, researchers began to explore alternatives that could overcome these issues.

A further limitation, shared in fact with standard CNNs and Transformers, is that vanilla RNNs assume a \emph{regular} sampling grid $(t_1, t_2, \ldots)$ with constant $\Delta t$. Many scientific time series violate this: ground-based photometric observations are interrupted by weather and the day/night cycle, space missions incur telemetry dropouts, and clinical records are updated event-by-event. A line of research has therefore reshaped the sequential architecture itself so that arbitrary inter-observation intervals $\Delta t_i$ enter natively. \emph{Neural ODEs} \cite{ChenNODE2018} replace the discrete hidden-state update $h_t = f(h_{t-1}, x_t)$ with a continuous-time evolution $dh/dt = g_\theta(h, t)$, so the state at any query time is obtained by numerical integration. Building on this idea, \emph{Latent ODE} and ODE-RNN models \cite{Rubanova2019} use an ODE as the decoder of an encoder--decoder architecture, producing predictions at arbitrary times. At the input level, \emph{GRU-D} \cite{Che2018} augments a GRU with learnable exponential-decay terms so that the hidden state explicitly ages during gaps and missingness is treated as a first-class feature rather than an inconvenience. The same goal can be pursued on the attention side through time-aware positional encodings and continuous-time attention variants, which we touch on in Section~\ref{sec:transformers}. These architectures are directly relevant to astronomical light curves and are revisited in Section~\ref{sec:outlook}: ZTF and LSST-era surveys will generate sequences where irregular cadence is the rule, not the exception.

\subsection{Convolutional Networks for Time Series}\label{sec:cnn}
Convolutional neural networks are well-known for image and signal processing tasks, but they can also be applied to one-dimensional time series. A 1D CNN slides convolutional filters over the sequence to detect local patterns. Initially, one might think a CNN is only good for spatially local features, but with techniques like \textit{dilated convolutions} and deep stacking, CNNs can actually capture long-range dependencies as well. An advantage of CNNs is that they do not have an intrinsic sequential dependency – thus, they can be parallelized and are often faster to train than RNNs. They also have fewer issues with vanishing gradients due to their feed-forward nature.

One influential result by Bai et al. (2018) showed that a Temporal Convolutional Network (TCN), which is a 1D CNN architecture with causal convolutions (no leaking from future to past) and dilations (skipping inputs to exponentially increase receptive field), outperformed canonical RNNs like LSTMs on a suite of sequence modeling tasks. The TCN was able to achieve longer effective memory and better accuracy, suggesting that convolutional architectures can serve as a viable alternative to RNNs for many problems \cite{Bai2018}. The key aspects of the TCN are: (1) Causal convolutions ensure the model is feed-forward in time (outputs at time $t$ only depend on inputs up to $t$), and (2) Dilated convolutions allow the receptive field to grow exponentially with depth, meaning a relatively shallow network can model very long sequences. Residual connections are also used to ease training of deep networks. After this work, CNN-based models have gained traction in time series. For instance, in healthcare time series such as patient vital-sign monitoring, CNNs have been used to detect patterns of disease onset. In traffic flow forecasting or electricity load forecasting, CNNs -- sometimes in combination with RNNs -- have achieved strong results by capturing local trends and repeating patterns.

Another example blending CNNs and time series is the WaveNet architecture \cite{Oord2016} (originally developed for audio waveform generation by DeepMind). WaveNet is essentially a dilated CNN that generates audio samples and has been adapted for other time series forecasting with good success. CNNs are also widely used in computer vision approaches to time series, such as treating spectrograms or recurrence plots of time series as images and then using 2D CNNs; however, those stray from the pure time-domain modeling.

In astronomy, convolutional networks have proven particularly effective. George \& Huerta (2018) demonstrated that 1D CNNs applied directly to LIGO strain time series can detect gravitational-wave signals from compact binary coalescences with accuracy comparable to matched filtering, but orders of magnitude faster \cite{George2018} -- a result with significant implications for real-time detection pipelines. CNNs have also been employed for classifying periodic variable star light curves by learning shape features automatically, and for real-time transient detection in survey data streams. The success of convolutional approaches shows that explicitly sequential processing (like RNN) is not the only way to model sequences -- learnable filters can capture temporal structure efficiently, and their parallelizability makes them well-suited to the high data rates of modern astronomical surveys.

\subsection{Attention Mechanisms and Transformers}\label{sec:transformers}
\begin{figure}[t]
\centering
\begin{tikzpicture}[
    node distance=0.9cm and 1.5cm,
    every node/.style={font=\small},
    block/.style={draw, rounded corners, minimum width=1.8cm, minimum height=0.6cm,
                  align=center, fill=white},
    arr/.style={-{Stealth[length=2mm]}, thick}
]
\node[block, fill=gray!10]                    (i1) {$x_1$};
\node[block, fill=gray!10, right=0.5cm of i1] (i2) {$x_2$};
\node[block, fill=gray!10, right=0.5cm of i2] (i3) {$x_3$};
\node[block, fill=gray!10, right=0.5cm of i3] (i4) {$x_T$};

\node[block, above=0.6cm of i1] (p1) {$+$ PE};
\node[block, above=0.6cm of i2] (p2) {$+$ PE};
\node[block, above=0.6cm of i3] (p3) {$+$ PE};
\node[block, above=0.6cm of i4] (p4) {$+$ PE};
\draw[arr] (i1) -- (p1);
\draw[arr] (i2) -- (p2);
\draw[arr] (i3) -- (p3);
\draw[arr] (i4) -- (p4);

\coordinate (upperbase) at ($(p1.north)!0.5!(p4.north)$);
\node[block, above=1.0cm of upperbase, anchor=south, minimum width=8.4cm,
      fill=gray!30, font=\bfseries\small] (sa) {Multi-Head Self-Attention};
\draw[arr] (p1.north) -- (p1.north |- sa.south);
\draw[arr] (p2.north) -- (p2.north |- sa.south);
\draw[arr] (p3.north) -- (p3.north |- sa.south);
\draw[arr] (p4.north) -- (p4.north |- sa.south);

\node[block, above=0.9cm of sa, minimum width=8.4cm,
      fill=gray!15] (ffn) {Feed-Forward Network};
\draw[arr] (p1.north |- sa.north) -- (p1.north |- ffn.south);
\draw[arr] (p2.north |- sa.north) -- (p2.north |- ffn.south);
\draw[arr] (p3.north |- sa.north) -- (p3.north |- ffn.south);
\draw[arr] (p4.north |- sa.north) -- (p4.north |- ffn.south);

\node[block, above=0.9cm of ffn, minimum width=8.4cm] (out) {Output Representations};
\draw[arr] (p1.north |- ffn.north) -- (p1.north |- out.south);
\draw[arr] (p2.north |- ffn.north) -- (p2.north |- out.south);
\draw[arr] (p3.north |- ffn.north) -- (p3.north |- out.south);
\draw[arr] (p4.north |- ffn.north) -- (p4.north |- out.south);

\coordinate (tb@top) at ([xshift=-0.30cm]ffn.north west);
\coordinate (tb@bot) at ([xshift=-0.30cm]sa.south west);
\draw[thick] (tb@top) -- (tb@bot);
\draw[thick] (tb@top) -- ++(0.12cm, 0);
\draw[thick] (tb@bot) -- ++(0.12cm, 0);
\node[left=6pt, rotate=90, font=\small] at ($ (tb@top)!0.5!(tb@bot) $) {Transformer Block};
\end{tikzpicture}
\caption{Simplified architecture of a Transformer encoder for time series. The input sequence $(x_1, \ldots, x_T)$ is augmented with positional encodings (PE) to preserve temporal order. The self-attention mechanism allows each time step to attend to all others, capturing long-range dependencies. The feed-forward network is applied token-wise, preserving the $T$ parallel streams up to the output. Multiple such blocks can be stacked for greater representational power.}
\label{fig:transformer}
\end{figure}
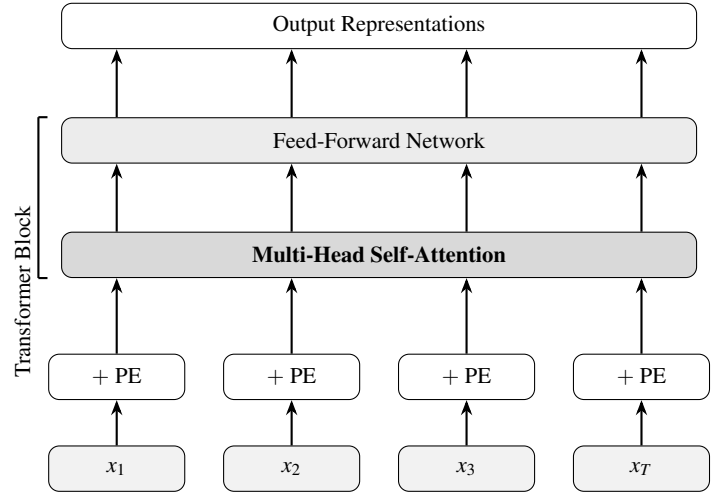

The latest development in sequence modeling has been the rise of \textit{Transformers}, which forgo both recurrence and convolution in favor of a mechanism called \textit{self-attention}. The Transformer architecture, introduced by Vaswani et al. (2017) in the context of natural language processing, has since become state-of-the-art in Natural Language Processing (NLP) and is making inroads in time series analysis as well \cite{Vaswani2017}. The core idea of self-attention is to allow each element of a sequence to attend to (i.e., compute weighted interactions with) every other element, capturing long-range dependencies directly. Given an input sequence, each element is projected into three vectors -- a \emph{query} $\mathbf{q}$, a \emph{key} $\mathbf{k}$, and a \emph{value} $\mathbf{v}$ -- and the attention output is computed as:
\begin{equation}
\text{Attention}(\mathbf{Q}, \mathbf{K}, \mathbf{V}) = \text{softmax}\!\left(\frac{\mathbf{Q}\mathbf{K}^\top}{\sqrt{d_k}}\right)\mathbf{V} \,,
\label{eq:attention}
\end{equation}
where $d_k$ is the dimension of the key vectors and the softmax ensures that the attention weights sum to one. The scaling factor $\sqrt{d_k}$ prevents the dot products from becoming too large, which would push the softmax into regions of very small gradients. In a Transformer model, the sequence is processed as a whole (not one step at a time as in an RNN), and multiple attention heads operate in parallel to capture different types of relationships (see Fig.~\ref{fig:transformer} for a schematic). The position in the sequence is encoded via positional encodings since the model itself has no inherent notion of order. For irregularly sampled series, time-aware positional encodings and continuous-time attention variants (see Section~\ref{sec:rnn}) extend these models to non-uniform grids.

For time series forecasting, Transformers offer the appealing ability to look at very long input histories and identify which past time points are most relevant to predicting the future. This can be very useful for data with long-term seasonal effects or irregular but long-range dependencies. There have been numerous Transformer-based models proposed for time series, often with modifications to handle time series challenges like very long sequences, high sampling frequency, or multiple correlated variables. Some notable examples include the Time Series Transformer for forecasting, the LogTrans and Informer models \cite{Zhou2021} which introduce sparse attention to handle long sequences efficiently, and the Temporal Fusion Transformer \cite{Lim2021} which integrates static and dynamic features with an attention mechanism for interpretable forecasting.

Transformers have been successfully applied to a variety of time series tasks, including univariate and multivariate forecasting, anomaly detection (where a Transformer can learn normal patterns and identify deviant behavior), and classification (mapping an entire sequence to a class, useful in medicine for classifying EEG signals, or in astronomy for classifying light curves). A comprehensive survey by Wen et al.\ (2023) categorizes time series Transformers and notes that many incorporate seasonality models or combine with recurrent/convolutional modules to better handle local patterns \cite{Wen2023}. One challenge with Transformers is that they can require very large amounts of data to train (they have many parameters and do not encode as much prior structure as, say, an AR model). They also historically have quadratic complexity in sequence length (due to attention over all pairs of points), but new architectures strive to reduce that, enabling longer sequences (e.g., the Informer uses probabilistic sampling to achieve linear complexity). A notable variant is PatchTST \cite{Nie2023}, which segments the input time series into patches (analogous to image patches in Vision Transformers) before feeding them into the Transformer encoder. This reduces the effective sequence length and allows the model to capture both local semantic information within patches and global dependencies across them, achieving state-of-the-art results on several long-term forecasting benchmarks.

It is also worth mentioning \textit{N-BEATS} (Neural Basis Expansion Analysis for Time Series) \cite{Oreshkin2020}, a pure deep learning architecture for forecasting that does not use recurrence, convolution, or attention. Instead, N-BEATS uses a stack of fully connected layers organized into blocks that learn basis expansion coefficients, producing both a forecast and a backcast at each block. N-BEATS achieved top results in the M4 forecasting competition \cite{Makridakis2020M4}, demonstrating that carefully designed feed-forward architectures can be highly competitive.

In practice, Transformers and these specialized architectures have begun to outperform RNNs on long-horizon forecasting tasks and remain an active area of research.

In practice, hybrid models combining the above approaches often yield the best results. For example, one might use a CNN or RNN to encode local structure and then an attention layer to select important time steps (this is seen in some medical time series models). Or use an RNN to get a first pass prediction and a Transformer to adjust for long-term corrections.

Deep learning models, while powerful, also bring challenges: they are often less interpretable than statistical models (though attention weights can sometimes be interpreted, and there is work on explaining neural forecasts), and they require careful tuning and significantly more data to avoid overfitting. In scientific domains like astrophysics, where understanding the model can be as important as predictive accuracy, this is an important consideration. Researchers often employ techniques like saliency or integrated gradients to interpret what an RNN or CNN might be focusing on in a light curve, or use attention weights in a Transformer to see which past events influenced a supernova classification decision. However, whether attention weights constitute a faithful explanation of model decisions has been debated \cite{JainWallace2019, WiegreffePinter2019}.

Finally, beyond supervised learning, deep learning enables new paradigms for time series: for instance, \textit{generative models} such as Variational Autoencoders (VAEs) and Generative Adversarial Networks (GANs) have been used to generate synthetic time series data or to impute missing data. In cosmology, one might train a VAE on simulated time series of certain phenomena and use it to generate more examples for data augmentation. Or use a GAN to fill in gaps in a sparsely sampled signal.

\subsection{Foundation Models for Time Series}
A rapidly emerging trend is the development of \textit{foundation models} for time series---large, pre-trained models that can generalize across diverse tasks and domains with minimal or no task-specific fine-tuning. Inspired by the success of Large Language Models (LLMs) in natural language processing, several groups have recently proposed pre-trained time series models. Notable examples include TimesFM \cite{Das2024}, a decoder-only Transformer pre-trained on a large corpus of real-world and synthetic time series, and Chronos \cite{Ansari2024}, which tokenizes time series values and trains a language model architecture on them. These models can perform zero-shot forecasting on unseen time series, often achieving competitive accuracy with task-specific models that were trained on the target data. Other approaches, such as Lag-Llama \cite{Rasul2024}, adapt existing LLM architectures for probabilistic time series forecasting using lags as covariates.

The appeal of foundation models lies in their potential to democratize time series analysis: a single pre-trained model could be applied out-of-the-box to forecasting tasks across astronomy, finance, healthcare, and engineering without requiring domain-specific training data or feature engineering. However, important open questions remain regarding their robustness to distribution shifts, their ability to handle highly irregular or multivariate scientific data, and whether they can match the performance of carefully tuned domain-specific models. In astrophysics, where data have unique noise properties and cadences, the applicability of these general-purpose models is an active area of investigation.

\subsection{Deep Learning for Time Series Classification}\label{sec:dl-tsc}
Beyond forecasting, time series classification (TSC) is a second major problem tackled by deep learning. The feature-based and ROCKET approaches of Section~\ref{sec:feature-based} already provide strong, computationally cheap baselines; neural classifiers aim to match or exceed them by learning features end-to-end.

Early deep baselines for TSC include Fully Convolutional Networks (FCN) and 1-D ResNets, which stack 1-D convolutions followed by global average pooling before a softmax output. \textit{InceptionTime} \cite{IsmailFawaz2020}, an ensemble of 1-D Inception-style networks, was the first neural architecture to match HIVE-COTE on the UCR/UEA archives and has since become a de-facto deep-learning benchmark for TSC. \textit{HIVE-COTE 2.0} \cite{Middlehurst2021} remains a leading non-deep reference -- a heterogeneous meta-ensemble fusing shape-, interval-, dictionary-, and deep-based classifiers; for streaming astronomical pipelines InceptionTime and MiniRocket (Section~\ref{sec:feature-based}) are typically preferred because HIVE-COTE 2.0 is accurate but expensive.

Transformer-based classifiers have followed the general trend. The Time Series Transformer of Zerveas et al.\ \cite{Zerveas2021} pre-trains a multivariate Transformer encoder with a masked-value objective and then fine-tunes for classification, regression, or imputation, achieving accuracy competitive with InceptionTime on UEA benchmarks and offering substantial gains when large unlabelled datasets are available for pre-training.

A natural astronomical target is the classification of variable-star light curves from survey streams such as ZTF and LSST, where per-object cadence is irregular and class distributions are strongly imbalanced; these deep classifiers underlie several current pipelines, as illustrated by the variable-star classification case study in Section~\ref{sec:examples}.

\subsection{Comparative Overview}
Table~\ref{tab:comparison} provides a comparative summary of the main method families discussed in this chapter. The choice of method depends on the specific task (forecasting, classification, anomaly detection), the data characteristics (size, regularity, dimensionality), and practical constraints (computational budget, interpretability requirements). As a general guideline: for small to moderate datasets with strong seasonal or autoregressive structure, classical methods (ARIMA, exponential smoothing) remain highly competitive and should serve as baselines. For tabular or feature-engineered time series data, gradient-boosted trees (XGBoost, LightGBM) are often the best starting point. For large datasets with complex patterns, deep learning methods offer the greatest flexibility, with Transformers and their variants currently representing the frontier for long-horizon forecasting.

\begin{table}[t]
\centering
\caption{Comparative summary of time series analysis methods. Computational cost refers to training; $N$ is the number of observations, $T$ the sequence length.}
\label{tab:comparison}
\small
\begin{tabular}{p{2.2cm}p{2.4cm}p{2.4cm}p{1.8cm}cp{1.6cm}}
\hline
\textbf{Method} & \textbf{Strengths} & \textbf{Limitations} & \textbf{Best for} & \textbf{Astro-native}\textsuperscript{\ddag} & \textbf{Cost} \\
\hline
ARIMA / SARIMA & Interpretable, fast, well-understood theory & Linear, univariate, requires stationarity & Short-term forecasting & $\times$ & $\mathcal{O}(N)$ \\
Exp.\ Smoothing & Robust, minimal tuning & Limited to trend + season & Seasonal forecasting & $\times$ & $\mathcal{O}(N)$ \\
Kalman Filter & Optimal for linear-Gaussian, recursive & Linear assumption & Tracking, filtering & \checkmark & $\mathcal{O}(N)$ \\
CARMA / DRW & Continuous-time; closed-form likelihood at irregular $t_i$ & Gaussian driving noise; stationarity & Quasar/AGN variability & \checkmark & $\mathcal{O}(N)$ \\
Gaussian Processes & Flexible, principled UQ, handles irregular data & $\mathcal{O}(N^3)$ naive; kernel choice & Regression with UQ & \checkmark & $\mathcal{O}(N^3)$* \\
Tree Ensembles & Nonlinear, robust, fast, handles mixed features & No native sequential structure & Tabular / feature-based & partial & $\mathcal{O}(N \log N)$ \\
ROCKET & Extremely fast, near-SOTA classification & Classification only & TS classification & partial & $\mathcal{O}(N)$ \\
RNN / LSTM & Learns sequential patterns, flexible & Slow training, vanishing gradients & Seq-to-seq tasks & partial & $\mathcal{O}(N T)$ \\
TCN & Parallelizable, long memory via dilation & Fixed receptive field & Forecasting, generation & partial & $\mathcal{O}(N T)$ \\
Transformer & Global attention, long-range deps. & Data-hungry, $\mathcal{O}(T^2)$ & Long-horizon, multivariate & partial & $\mathcal{O}(N T^2)$** \\
Foundation Models & Zero-shot, general-purpose & Robustness unclear, large & Quick baseline & $\times$\textsuperscript{\dag} & Inference only \\
\hline
\multicolumn{6}{l}{\footnotesize *$\mathcal{O}(N)$ with sparse approximations (e.g.\ \texttt{celerite}). **$\mathcal{O}(NT\log T)$ with efficient variants (Informer).}\\
\multicolumn{6}{l}{\footnotesize \textsuperscript{\dag}Unclear robustness to irregular sampling and astronomical noise properties.}\\
\multicolumn{6}{l}{\footnotesize \textsuperscript{\ddag}Astro-native: natively handles irregular sampling and/or per-point observational uncertainties.}
\end{tabular}
\end{table}

\section{Examples and Case Studies}\label{sec:examples}
To illustrate how these concepts come together, we briefly consider a few cross-domain examples.

\paragraph{Variable Star Classification (Astrophysics).} Astronomers often need to classify a star as a Cepheid, RR Lyrae, eclipsing binary, etc., based on its brightness over time. Historically, this was done by extracting features like period, amplitude, and certain specific metrics from the light curve and feeding into classifiers. Today, one can apply a deep learning model: e.g., an LSTM or Transformer that reads the raw light curve (time, flux values) and outputs a class probability. The model might implicitly learn to detect periodic patterns or characteristic shapes (like the sawtooth waveform of a classical Cepheid). Akhmetali et al. (2025) provide a survey of ML in light curve analysis, highlighting how these modern approaches outperform traditional methods and noting challenges like class imbalance and noise \cite{Akhmetali2025}. For instance, the paper discusses how a convolutional neural network can automatically learn features corresponding to astrophysical properties and how semi-supervised learning can leverage unlabeled data to improve classification \cite{Akhmetali2025}.

\paragraph{Macroeconomic Forecasting (Economics).} Predicting indicators like inflation or unemployment often uses ensembles of approaches. A classical ARIMA or vector autoregression might capture short-term dynamics, while a machine learning model (e.g., XGBoost with many features such as commodity prices, stock indices, etc.) adds predictive power by capturing nonlinear effects and interactions. Recently, deep models like sequence-to-sequence LSTMs have been explored to jointly forecast multiple economic time series. These models can learn complex cross-correlations (for example, how an interest rate change might influence exchange rates over time). However, given the relatively small datasets in macroeconomics (quarterly data points measured over decades), careful regularization and often hybrid approaches with econometric models are used to avoid overfitting.

\paragraph{Weather and Climate (Environmental Science).} Weather prediction is traditionally done with physics-based models (numerical weather prediction), but machine learning has started to play a role in post-processing and even direct forecasting of certain variables. For example, ARIMA models have been used for local weather variable forecasting as a baseline, but now LSTM and CNN-based models can capture nonlinear relationships like how temperature, humidity, wind speed time series interact. In the context of climate science, time series of global temperatures have trends that ARIMA can model to some extent (with integration), but ML might be used to combine multiple correlated series (solar cycles, ocean indices) for improved predictions. For instance, a GRU network can be trained to predict precipitation for the next day at a station given not only that station’s past readings but also neighboring stations’ readings (a multivariate sequence problem). The flexibility of ML allows incorporating many data sources (satellite data as additional features, etc.) that classical univariate models would ignore.

\paragraph{Anomaly Detection in Manufacturing (Industry).} Consider a factory sensor that records pressure readings over time. A sudden deviation might indicate a malfunction. Using an unsupervised LSTM autoencoder, one can train on a period of normal operations. The LSTM learns to reconstruct the normal time series patterns. When a new sequence comes in, it is fed through the autoencoder; if the reconstruction error exceeds a threshold, the sequence is flagged as anomalous. Alternatively, a forecasting model (like a neural network that predicts the next value) can be used: if the actual next value falls outside a prediction interval, raise an alarm. These ML-based anomaly detection systems adapt to the data and can often detect subtle changes that rule-based thresholds would miss.

\paragraph{Gravitational Wave Detection (Cosmology).} The detection of gravitational waves by LIGO and Virgo represents one of the most spectacular triumphs of modern physics. The raw detector output is a continuous time series of strain data, buried in complex, non-stationary instrumental noise. Identifying a faint gravitational wave signal (lasting from milliseconds to minutes, depending on the source type) within this noisy stream is fundamentally a time series analysis problem. Matched filtering, a classical technique, cross-correlates the data with a bank of theoretical waveform templates, but is computationally expensive and limited to known signal morphologies. Deep learning approaches, particularly 1D CNNs and ResNets applied directly to the raw strain time series, have demonstrated the ability to detect compact binary coalescence signals with accuracy comparable to matched filtering, but at a fraction of the computational cost \cite{George2018}. This makes them promising candidates for real-time detection in future observing runs. Furthermore, for signals with poorly known waveforms (e.g., from core-collapse supernovae), ML-based anomaly detection on the time series offers a model-agnostic path to discovery.

\section{Software and Practical Tools}
A reference chapter would be incomplete without pointing the practitioner to available software implementations. Table~\ref{tab:software} summarizes the major open-source libraries for time series analysis, organized by category. The Python ecosystem is particularly rich: \texttt{statsmodels} provides classical methods (ARIMA, state-space models, exponential smoothing), while \texttt{scikit-learn} \cite{Pedregosa2011} offers a broad ML toolkit that can be applied to time series via feature engineering. For astronomical time series, \texttt{astropy} includes Lomb-Scargle periodogram implementations and \texttt{lightkurve} provides specialized tools for Kepler/TESS light curve analysis. On the deep learning side, \texttt{PyTorch Forecasting}, \texttt{GluonTS} \cite{Alexandrov2020}, and \texttt{Darts} offer high-level APIs for training and evaluating neural forecasting models including Transformers, N-BEATS, and temporal fusion transformers. For time series classification, \texttt{tslearn}, \texttt{sktime}, and \texttt{aeon} provide DTW, ROCKET, shapelet, and other algorithms in a unified interface. The \texttt{tsfresh} library automates feature extraction, while \texttt{Prophet} \cite{Taylor2018} offers an accessible interface for forecasting with trend, seasonality, and holiday effects.

\begin{table}[t]
\centering
\caption{Selected open-source software libraries for time series analysis. All listed libraries are available in Python unless otherwise noted.}
\label{tab:software}
\small
\begin{tabular}{lll}
\hline
\textbf{Category} & \textbf{Library} & \textbf{Key capabilities} \\
\hline
Classical methods & \texttt{statsmodels} & ARIMA, SARIMAX, state-space, exp.\ smoothing \\
 & \texttt{Prophet} & Trend + seasonality decomposition \\
 & \texttt{astropy} & Lomb-Scargle periodogram \\
\hline
Feature extraction & \texttt{tsfresh} & Automated extraction \& selection \\
 & \texttt{cesium} & Astronomical time series features \\
\hline
ML / Classification & \texttt{scikit-learn} & General ML (RF, GBM, SVM, \ldots) \\
 & \texttt{sktime} / \texttt{aeon} & TS classification, ROCKET, shapelets \\
 & \texttt{tslearn} & DTW, clustering, shapelets \\
\hline
Deep learning & \texttt{PyTorch Forecasting} & TFT, N-BEATS, DeepAR \\
 & \texttt{GluonTS} & Probabilistic forecasting models \\
 & \texttt{Darts} & Unified API for many DL models \\
\hline
Astronomy-specific & \texttt{lightkurve} & Kepler/TESS light curve analysis \\
 & \texttt{celerite2} & Fast GP for stellar variability \\
\hline
\end{tabular}
\end{table}

\section{Outlook and Challenges}\label{sec:outlook}
Time series analysis with machine learning has seen tremendous progress, but there remain important challenges and open issues, especially in scientific contexts:

\begin{itemize}
    \item \textbf{Interpretability:} As noted, methods like deep neural networks can be ``black boxes’’. In astrophysics and cosmology, scientists need interpretable models to trust discoveries. Efforts like attention mechanisms (highlighting which time steps influenced a decision), or hybrid models that combine neural nets with physics-based components, are ongoing to ensure interpretability. For example, in classifying an astrophysical transient, one might want the model to indicate which part of the light curve (e.g., the initial rise vs.\ the decline) was most indicative of it being a supernova versus a flare.

    \item \textbf{Irregular Sampling and Missing Data:} Many scientific time series are not sampled at fixed intervals (astronomical observations might have gaps due to weather or satellite orbits). Classical methods like ARIMA can be extended with missing data techniques, and Gaussian processes naturally handle irregular times. But many ML and DL models assume regular time steps. New architectures or preprocessing (like interpolation or using time as an explicit input to networks) are needed. In cosmology, one might have data at non-uniform cosmic time points -- handling that in a neural network requires care.

    \item \textbf{Heteroscedastic Noise and Per-point Uncertainties:}\label{item:hetero} Astronomical time series -- and many scientific time series more broadly -- arrive with known per-observation uncertainties $\sigma_i$, from photometric calibration, detector readout noise, or propagated measurement errors. This information is a strong prior that generic ML pipelines routinely discard. Gaussian processes and \texttt{celerite}-style state-space models incorporate it natively through the noise term $\sigma_n^2\,\mathbf{I}$ of Eq.~\eqref{eq:gp_mean}, in practice replaced by an inhomogeneous diagonal $\mathrm{diag}(\sigma_i^2)$. Deep models can recover this capability in three main ways: (i) weighted losses, e.g.\ a weighted mean-squared error with $w_i = 1/\sigma_i^2$; (ii) Gaussian negative-log-likelihood heads that condition on the known $\sigma_i$; and (iii) fully probabilistic approaches -- deep ensembles, variational inference, or MC Dropout \cite{Gal2016} -- that additionally propagate \emph{model} uncertainty on top of the input noise. The same concern carries into evaluation: heteroscedastic data call for weighted or calibration-aware extensions of the metrics of Section~\ref{sec:evaluation-metrics} (e.g.\ weighted CRPS), because a constant-$\sigma$ baseline is misleading when the noise level varies by orders of magnitude across the light curve.

    \item \textbf{Long Sequence Lengths:} Some phenomena have very long-term dependencies (financial markets might have multi-year cycles, climate has multi-decadal oscillations). Models like Transformers are being developed to handle longer sequences (with memory or with hierarchical attention). Even so, there is often a trade-off between sequence length and model complexity. Techniques such as sequence chunking, or multiresolution networks (looking at the series at different time scales), are active research areas.

    \item \textbf{Small Data Regimes:} Not all time series problems have big data. In cosmology, we might only have one universe’s worth of data for certain signals -- we cannot get more samples. In such cases, incorporating domain knowledge (like physics equations) into ML models can help. There is also interest in Bayesian deep learning to better quantify uncertainty when data are limited.

    \item \textbf{Physics-Informed Machine Learning:} A particularly promising direction for scientific applications is the integration of physical laws directly into ML architectures. Physics-Informed Neural Networks (PINNs) \cite{Raissi2019} embed differential equations as soft constraints in the loss function, enabling the network to learn dynamics that are consistent with known physics even when data are sparse. A complementary direction is offered by neural operators such as FNO \cite{LiFNO2021} and DeepONet \cite{LuDeepONet2021}, which learn mappings between function spaces rather than pointwise solutions. For time series in astrophysics and cosmology -- where the underlying physical processes are often partially known (e.g., orbital mechanics, radiative transfer, cosmological expansion) -- physics-informed approaches can dramatically improve generalization and interpretability compared to purely data-driven models. Hybrid architectures that combine a physics-based ``backbone’’ with a neural network residual for unmodeled effects represent a middle ground between full physical modeling and black-box ML.

    \item \textbf{Evaluation and Benchmarks:} Time series forecasting is usually evaluated by error metrics (RMSE, MAPE, etc.), but in some cases (anomaly detection, etc.) it is less clear how to objectively measure success. Moreover, a model that is best on average may fail on critical rare cases. The ML community has started to create large benchmark datasets and competitions (like the M series and others) to drive progress. However, astronomical datasets have unique noise characteristics and evaluation needs (e.g., detecting a rare one-time event). Cross-pollination between domains (bringing methods from, say, econometrics to astronomy and vice versa) can be fruitful.
\end{itemize}

\section{Conclusions}
Time series analysis in machine learning is a rich field at the intersection of statistics, computer science, and domain sciences. Fundamental concepts like stationarity and autocorrelation continue to guide how we model temporal data, while ML has provided powerful new tools to capture complex behaviors that classical models alone cannot handle. Tree-based models and other ML algorithms offer strong predictive performance with comparatively modest engineering effort, while deep neural networks provide even greater flexibility by learning directly from raw data -- at the cost of higher complexity and data requirements. In astrophysics and cosmology, these techniques are enabling discoveries in the time domain, from identifying new classes of stellar variability to enhancing the analysis of gravitational-wave signals. By combining theoretical understanding of both the algorithms and the scientific problem, researchers can build models that not only predict well but also yield physical insight. As data volumes continue to grow -- the LSST alone is expected to produce an unprecedented number of time series -- mastering these ML techniques will be essential for transforming data into knowledge.

\begin{acknowledgement}
The authors thank the editors of this volume for the invitation to contribute this chapter.
\end{acknowledgement}


\end{document}